\documentclass[Physsubmission, Phys]{SciPost}

\hypersetup{
    colorlinks,
    linkcolor={red!50!black},
    citecolor={blue!50!black},
    urlcolor={blue!80!black}
}

\usepackage[bitstream-charter]{mathdesign}
\usepackage{graphics}
\graphicspath{{./figures/}}

\newcommand{\drv}{{\rm d}}

\newcommand{\tcite}[1]{~\cite{#1}}
\newcommand{\tref}[1]{~\ref{#1}}

\DeclareSymbolFont{usualmathcal}{OMS}{cmsy}{m}{n}
\DeclareSymbolFontAlphabet{\mathcal}{usualmathcal}

\begin{document}

\begin{center}{\Large \textbf{
Exclusive emissions of rho-mesons \\ and the unintegrated gluon distribution
}}\end{center}

\begin{center}
Andr\`ee Dafne Bolognino\textsuperscript{1,2},
Francesco Giovanni Celiberto\textsuperscript{3,4,5$\star$}, \\
Dmitry Yu. Ivanov\textsuperscript{6} and
Alessandro Papa\textsuperscript{1,2}
\end{center}

\begin{center}
{\bf 1} Dipartimento di Fisica, Universit\`a della Calabria, \\ I-87036 Arcavacata di Rende, Cosenza, Italy
\\
{\bf 2} Istituto Nazionale di Fisica Nucleare, Gruppo collegato di Cosenza, \\ I-87036 Arcavacata di Rende, Cosenza, Italy
\\
{\bf 3} European Centre for Theoretical Studies in Nuclear Physics and Related Areas (ECT*), I-38123 Villazzano, Trento, Italy
\\
{\bf 4} Fondazione Bruno Kessler (FBK), I-38123 Povo, Trento, Italy
\\
{\bf 5} INFN-TIFPA Trento Institute of Fundamental Physics and Applications, \\ I-38123 Povo, Trento, Italy
\\
{\bf 6} Sobolev Institute of Mathematics, 630090 Novosibirsk, Russia
\\
* fceliberto@ectstar.eu
\end{center}

\begin{center}
\today
\end{center}


\definecolor{palegray}{gray}{0.95}
\begin{center}
\colorbox{palegray}{
  \begin{tabular}{rr}
  \begin{minipage}{0.1\textwidth}
    \includegraphics[width=22mm]{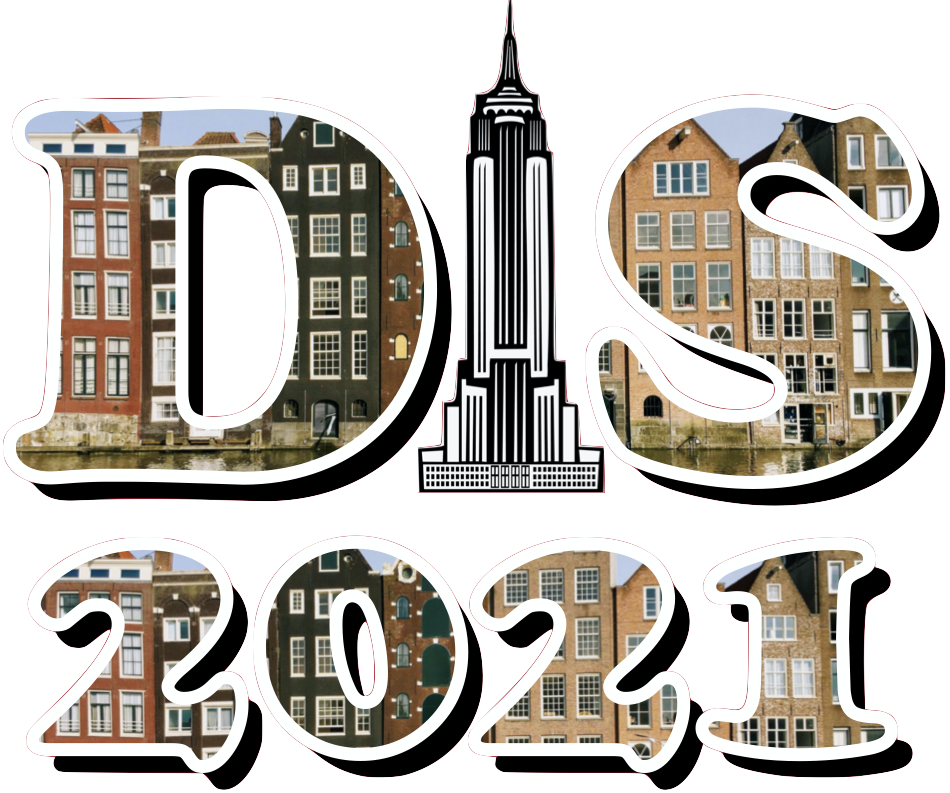}
  \end{minipage}
  &
  \begin{minipage}{0.75\textwidth}
    \begin{center}
    {\it Proceedings for the XXVIII International Workshop\\ on Deep-Inelastic Scattering and
Related Subjects,}\\
    {\it Stony Brook University, New York, USA, 12-16 April 2021} \\
    \doi{10.21468/SciPostPhysProc.?}\\
    \end{center}
  \end{minipage}
\end{tabular}
}
\end{center}

\section*{Abstract}
{\bf
Exclusive emissions of vector mesons in forward directions of rapidity offer us a faultless chance to probe the proton structure at small $x$. A high-energy factorization formula is established within BFKL, given as the convolution of an impact factor depicting the forward-meson emission and of an unintegrated gluon distribution (UGD) driving the gluon evolution at small $x$. As a nonperturbative quantity, the UGD is not well known and several models for it exist. We present recent progresses on the study of the exclusive forward $\rho$-meson leptoproduction at HERA and EIC energies, showing how osbervables sensitive to different polarization states of the $\rho$-particle act as discriminators for the existing UGD models.
}

\vspace{10pt}
\noindent\rule{\textwidth}{1pt}
\tableofcontents\thispagestyle{fancy}
\noindent\rule{\textwidth}{1pt}
\vspace{10pt}

\section{Introduction}
\label{sec:intro}

Inclusive as well as exclusive emissions of particles in forward directions of rapidity are widely recognized as excellent channels to probe the high-energy dynamics of strong interactions in new and original ways. The Balitsky--Fadin--Kuraev--Lipatov (BFKL)\tcite{Fadin:1975cb,Kuraev:1977fs,Balitsky:1978ic} resummation of energy logarithms provides us with a factorization formula of scattering amplitudes (and thence, for inclusive reactions, of cross sections) given as the convolution of a universal gluon Green's function and two process-dependent impact factors depicting the transition from each colliding particle to the corresponding final-state object.
The first class of reactions that allows us to test the BFKL mechanism in represented by the inclusive \emph{semi-hard} detection of two particles featuring large transverse momenta and well separated in rapidity. Here the established factorization is \emph{hybrid}, in the sense that the pure high-energy dynamics is supplemented by collinear distributions that enter expressions of impact factors. In the last three decades several phenomenological analyses have been proposed for distinct semi-hard configurations (see, \emph{e.g.}, Refs.\tcite{Ducloue:2013hia,Ducloue:2013bva,Caporale:2014gpa,Celiberto:2016ygs,Caporale:2015int,Caporale:2016xku,Caporale:2016zkc,Celiberto:2016hae,Celiberto:2017ptm,Celiberto:2017nyx,Bolognino:2019yls,Boussarie:2017oae,Bolognino:2018oth,Bolognino:2019yqj,Golec-Biernat:2018kem,Celiberto:2020tmb,Bolognino:2021mrc,Celiberto:2021dzy,Celiberto:2021fdp,Celiberto:2022dyf}), this allowing us to define BFKL-sensitive observables as well as discriminate between BFKL and fixed-order calculations\tcite{Celiberto:2015yba,Celiberto:2015mpa,Celiberto:2020wpk}.
The second class of probes for BFKL is given by single forward emissions, where the proton content is accessed via the so-called unintegrated gluon distribution (UGD), whose evolution at small $x$ is driven by the gluon Green's function. As a nonperturbative quantity, the UGD in not well known and several phenomenological models for it have been proposed so far. Golden channels to study the UGD are the deep inelastic scattering at small-$x$\tcite{Hentschinski:2012kr}, the single exclusive electro- or photoproduction of vector mesons\tcite{Besse:2013muy,Bolognino:2018rhb,Bolognino:2018mlw,Bolognino:2019bko,Bolognino:2019pba,Celiberto:2019slj,Bautista:2016xnp,Garcia:2019tne,Hentschinski:2020yfm,Bolognino:2022uty,Celiberto:2022fam,Hentschinski:2022xnd} at HERA, and the forward Drell--Yan process\tcite{Brzeminski:2016lwh,Motyka:2016lta,Celiberto:2018muu} at LHCb.

In this paper we extend the analysis done in Ref.\tcite{Bolognino:2018rhb}, by showing how the study of helicity-amplitude ratios for the exclusive leptoproduction of $\rho$ mesons permits us to discriminate among different UGD models at HERA and EIC energies.

\section{Exclusive $\rho$-meson leptoproduction at HERA and the EIC}
\label{sec:rho}

The process under investigation is the single exclusive production of a $\rho$ meson in $ep$ collisions via the subprocess
\begin{equation}
\label{eq:subprocess}
 \gamma^*_{\lambda_i} (Q^2) \, p \; \to \; \rho_{\lambda_f} p \;,
\end{equation}
where a photon with virtuality $Q^2$ and polarization $\lambda_i$ is absorbed by the proton and a $\rho$ particle with polarization $\lambda_f$ is emitted in the final state. The two helicity states $\lambda_{i,f}$ can take values $0$ (longitudinal) and $\pm 1$ (transverse).
In the high-energy regime one observes a strict semi-hard scale ordering, $W^2 \gg Q^2 \gg \Lambda^2_{\rm QCD}$ ($W$ is the subprocess center-of-mass energy), that leads to small $x$, $x = Q^2/W^2$. Here, a high-energy factorized expression for polarized amplitudes holds
\begin{equation}
\label{eq:ampltude}
 {\cal T}_{\lambda_i \lambda_f}(W^2, Q^2) = \frac{i W^2}{(2 \pi)^2} \int \frac{\drv^2 q}{(q^2)^2} \; \Phi^{\gamma^*_{\lambda_i} \to \rho_{\lambda_f}}(q^2, Q^2) \, {\cal F}_g (x, Q^2) \;,
\end{equation}
where $\Phi^{\gamma^*_{\lambda_i} \to \rho_{\lambda_f}}(q^2, Q^2)$ is the impact factor portraying the photon-to-$\rho$ transition that embodies distribution amplitudes (see Section~2.1 of Ref.\tcite{Bolognino:2018rhb} for details), and ${\cal F}_g (x, Q^2)$ is the UGD. We include in our analysis the six UGD models presented in Section~2.3 of Ref.\tcite{Bolognino:2018rhb}, together with the novel Bacchetta--Celiberto--Radici--Taels (BCRT) transverse-momentum dependent unpolarized gluon distribution\tcite{Bacchetta:2020vty,Celiberto:2021zww,Bacchetta:2021lvw,Bacchetta:2021twk,Bacchetta:2022esb}.

\begin{figure}[h]
\centering
\includegraphics[width=0.47\textwidth]{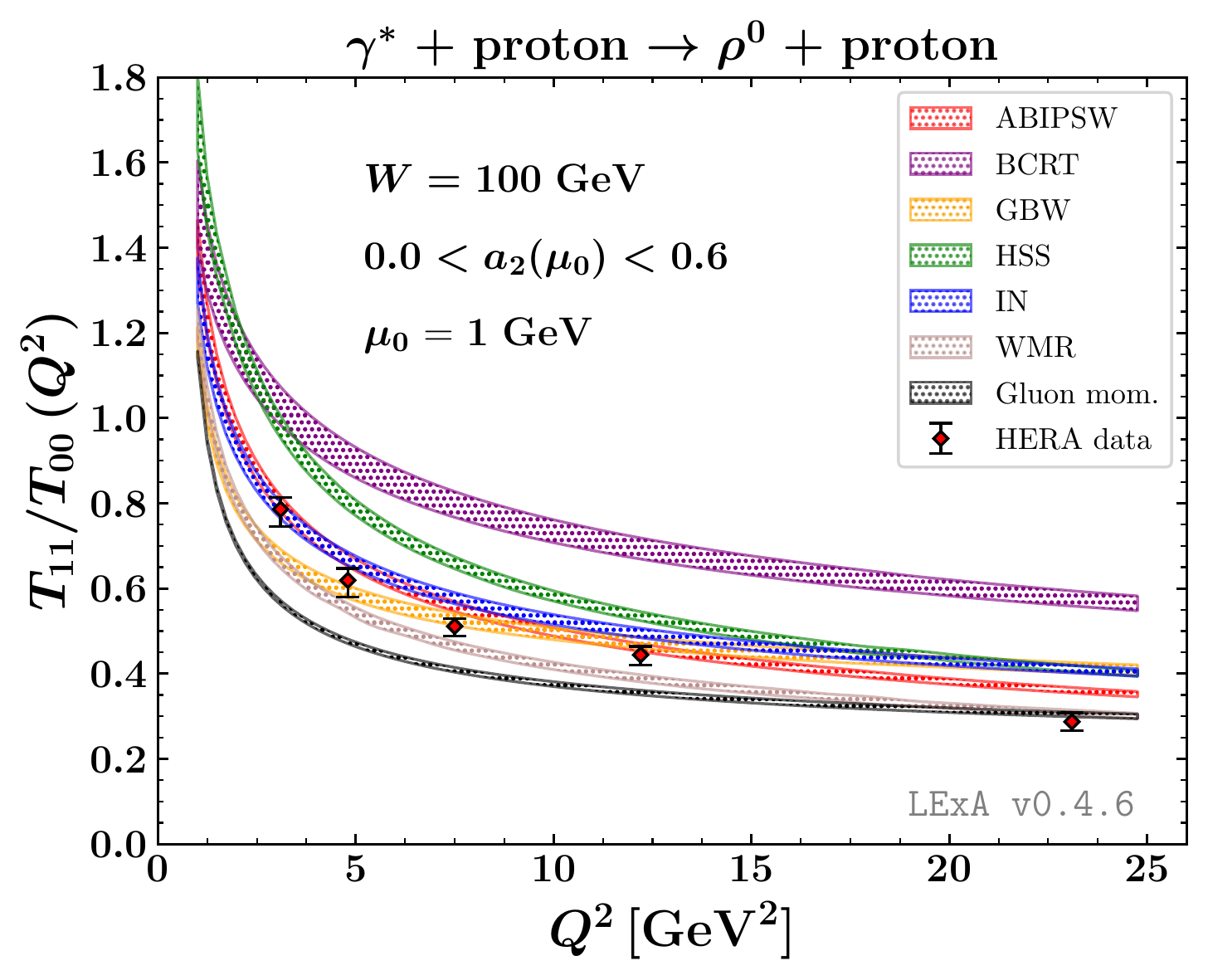} \hspace{0.5cm}
\includegraphics[width=0.47\textwidth]{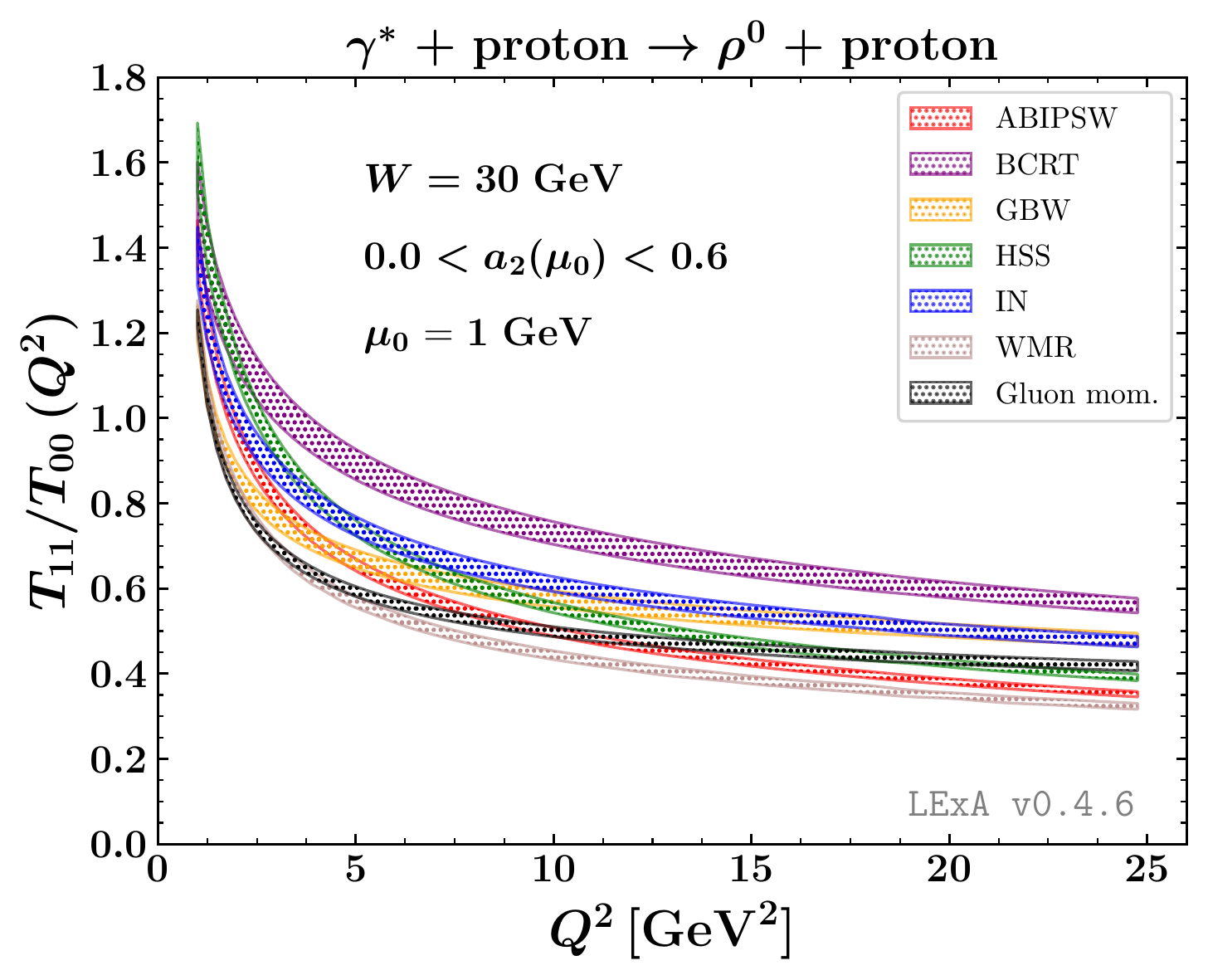}
\caption{$Q^2$-dependence of the polarized $T_{11}/T_{00}$ ratio, for all the considered UGD models, at $75$ GeV HERA (left) and at $30$ GeV EIC (right). 
Uncertainty bands represent the effect of varying $a_2(\mu_0 = 1\,$\rm GeV$)$ between $0.0$ and $0.6$.
Numerical results were done by using the \emph{Leptonic-Exclusive-Amplitudes} ({\tt LExA}) module as implemented in the {\tt JETHAD} environment\tcite{Celiberto:2020wpk}.}
\label{fig:rho}
\end{figure}

In Fig.\tref{fig:rho} we present our study for the ${\cal T}_{11} / {\cal T}_{00}$ helicity ratio as a function of $Q^2$. We compare our results with HERA data\tcite{Aaron:2009xp} at $W = 100$ GeV (left panel), and present new predictions for the EIC\tcite{AbdulKhalek:2021gbh,Bolognino:2021niq} at the reference energy of $W = 30$ GeV (right panel). We employed the twist-2 distribution amplitude for longitudinal impact factor and the full twist-3 one for the transverse case, and we assessed the impact of the distribution-amplitude collinear evolution by varying the nonperturbative parameter $a_2(\mu_0 = 1\,$\rm GeV$)$ in the range 0.0 to 0.6 (see Ref.\tcite{Bolognino:2018rhb}). We observe that our results are spread over a large window. None of the UGD models is in agreement with HERA data over the whole range of $Q^2$, although the ABIPSW, IN and GBW ones better catch the intermediate $Q^2$ range. Predictions at EIC energies exhibit a reduction of the distance between models, together with a change of their hierarchy for some regions of $Q^2$. All these features corroborate our statement that the ${\cal T}_{11} / {\cal T}_{00}$ ratio is potentially able to strongly constrain the small-$x$ UGD.

\section{Conclusions}
\label{sec:conclusions}

We presented results for the exclusive electroproduction of $\rho$ mesons in the high-energy limit of strong interactions, pointing out how ratios of polarized amplitudes are excellent observables to discriminate among several existing UGD models. Comparisons with HERA data clearly indicate that none of these model is able to reproduce the entire $Q^2$ spectrum. This outcome is confirmed also by predictions for the EIC, the distance between models being however less pronounced.

Future studies of the polarized production of vector mesons at new generation colliders will certainly accelerate progress of understanding of the proton structure at small $x$. In view of the current situation, we believe that a path towards the first extraction of the UGD from a global fit on data coming from the EIC\tcite{Accardi:2012qut,AbdulKhalek:2021gbh}, the HL-LHC\tcite{Chapon:2020heu}, the FPF\tcite{Anchordoqui:2021ghd,Feng:2022inv} and NICA-SPD\tcite{Arbuzov:2020cqg} has been now traced.

\bibliography{references.bib}

\nolinenumbers

\end{document}